# A Roadmap for Controlled and Efficient *n*-type Doping of Self-assisted GaAs Nanowires Grown by Molecular Beam Epitaxy

*Marta Orrù, Eva Repiso, Stefania Carapezzi, Alex Henning, Stefano Roddaro, Alfonso*

*Franciosi, Yossi Rosenwaks, Anna Cavallini, Faustino Martelli, and Silvia Rubini\**

M. Orrù, Dr. S. Rubini
IOM CNR Laboratorio TASC
S.S. 14 km 163.5
I-34149 Trieste, Italy
E-mail: rubini@iom.cnr.it

E. Repiso
IOM CNR Laboratorio TASC
S.S. 14 km 163.5
I-34149 Trieste, Italy
and
Università degli Studi di Trieste
Via Valerio 2
I-34127 Trieste, Italy

Dr. S. Carapezzi, Prof. Anna Cavallini
Department of Physics and Astronomy
Viale Berti Pichat 6/2
Bologna, I-40127, Italy

A. Henning, Prof. Y. Rosenwaks
Department of Physical Electronics
School of Electrical Engineering
Tel-Aviv University, Ramat-Aviv, Israel

Dr. S. Roddaro
NEST, Scuola Normale Superiore and Istituto Nanoscienze CNR
Piazza S. Silvestro 12
I-56127 Pisa, Italy
and
IOM CNR Laboratorio TASC
S.S. 14 km 163.5
I-34149 Trieste, Italy

Prof. A. Franciosi
IOM CNR Laboratorio TASC
S.S. 14 km 163.5
I-34149 Trieste, Italy
and
Università degli Studi di Trieste
Via Valerio 2
I-34127 Trieste, Italy
and

Elettra-Sincrotrone Trieste S.C.p.A.
Area Science Park, S.S. 14, Km 163.5
I-34149 Trieste, Italy

Dr. F. Martelli
IMM CNR,
via Fosso del Cavaliere 100
00133 Roma, Italy

*N*-type doping of GaAs nanowires has proven to be difficult because the amphoteric character of silicon impurities is enhanced by the nanowire growth mechanism and growth conditions. The controllable growth of *n*-type GaAs nanowires with carrier density as high as $10^{20}$ electron/cm$^3$ by self-assisted molecular beam epitaxy using Te donors is demonstrated here. Carrier density and electron mobility of highly doped nanowires are extracted through a combination of transport measurement and Kelvin probe force microscopy analysis in single-wire field-effect devices. Low-temperature photoluminescence is used to characterize the Te-doped nanowires over several orders of magnitude of the impurity concentration. The combined use of those techniques allows the precise definition of the growth conditions required for effective Te incorporation.

## 1. Introduction

The development of electronic devices based on semiconductor nanowires (NWs) requires careful control of carrier polarity and concentration. *N*-doping of GaAs epilayers grown by molecular beam epitaxy (MBE) is reliably and efficiently obtained through the incorporation of Si atoms that behave as donors. In GaAs NWs grown by MBE however, silicon impurities have found to exhibit a marked amphoteric character as a result of the NW growth mechanism and/or growth conditions.[1-3] This has stimulated the search for a reliable and efficient way to induce *n*-type doping in GaAs NWs. This is a technologically relevant issue, since a precise control of doping polarity and profile is a prerequisite for development of nanowire-based electronic devices, (see for instance Refs. 4,5). Among the various possible donor impurities for GaAs NWs, tellurium represents a good candidate since it is a very effective dopant in GaAs epilayers[6] and does not present any risk of amphoteric behavior. The available experimental evidence concerning the use of Te dopants in GaAs NWs is, however, still limited to few studies. *N-p* core-shell GaAs NWs showing photovoltaic effect have been obtained using Te as donor impurities during gas-source MBE, NW growth being induced by Au particles.[7,8] Au-induced growth of GaAs NWs by means of metal-organic vapor-phase-epitaxy using Te as a dopant[9,10] yielded an estimated *n*-type carrier concentration of $9 \times 10^{17}$ cm$^{-3}$. Recently self-assisted growth of GaAs NWs by means of solid source MBE in presence of a GaTe flux lead to changes in the Raman spectra as a function of the GaTe flux, which were interpreted as evidence of Te incorporation within the NW body.[11]

All those cited works are occasional attempts to use Te as donor, but despite the high potential of Te as such in GaAs NWs, to date no information about the real efficiency of its use as dopant in NWs, doping limit, carrier mobility and optical properties is available.

Not only that, in general quantitative experimental evidence about effective carrier densities is relatively limited in the NWs literature. Often this information is derived from transport measurements using approximate methods based on bulk related quantities.[9,12,13] Only recently the possibility of perform Hall measurements on single NWs has been demonstrated[14,15] and systematically applied to characterize the carrier concentration in Sn doped InP NWs.[16] Unfortunately this approach requires a complex device design and nanofabrication process. Field-effect based single wire measurements are relatively simple to implement and provide reliable estimate of carrier concentration although care should be taken due to possible electrode screening, contact resistance and surface-related effects.[17,18] In most cases field-effect based single-wire measurements allow comparisons between different nanostructures and the establishment of systematic correlation between growth conditions and carrier densities.

In this work we provide evidence of efficient n-type doping in GaAs NWs using Te as donor impurities. Thanks to the combination of conventional field-effect transport analysis, Kelvin probe microscopy and photoluminescence (PL) spectroscopy we demonstrate that very high electron densities ($\geq 10^{20}$ cm$^{-3}$) could be reproducibly obtained. The NWs were obtained by self-assisted MBE growth. We used solid source MBE and co-evaporation from a GaTe source during growth. Although Te is well known to be volatile, we demonstrate highly efficient doping in NWs grown at 580 °C and up to 640°C, i.e. at temperatures well above those used to dope epitaxial layers (535 °C or lower).[6,19,20] Higher growth temperatures allow the achievement of better electrical and optical properties, thanks to the reduction in the residual impurity incorporation.

We investigated the dependence of the incorporation of active Te impurities on GaTe flux and on the growth temperature ($T_g$) by measuring the carrier density by transport measurement in a NW field-effect transistor (NW-FET) configuration. Our investigation indicates that the best compromise between a good NW morphology and a high electron density is obtained for growth temperature of 610 °C. The carrier density and mobility estimated by field-effect transport measurements were cross-correlated with results obtained by Kelvin-probe force microscopy (KPFM). The combined analysis allowed us to isolate the impact of the contact resistance in our FET devices, which can lead to uncontrolled systematic errors in the estimate of the carrier mobility. In addition, the effect of Te incorporation was also examined by systematic low temperature photoluminescence (PL) studies. The results provided a semi-quantitative, and independent control of the doping level estimates throughout the Te flux range examined and allowed us to detect Te incorporation already at the lowest doping levels, at which transport measurements are challenging, due to the difficulty to fabricate reliable ohmic contacts on GaAs NWs. The optical fingerprint of Te doping obtained by PL becomes then a useful tool for quick evaluations of Te doping of GaAs NWs.

## 2. Nanowire growth and morphology

GaAs NWs were grown by self-assisted, solid-source MBE. The substrate temperature ($T_g$) was set at 580, 610 and 640°C. Four series of NW samples were grown at each temperature for different fluxes from a GaTe Knudsen cell (see Section 7 for more details about the growth). The series are indexed according to the electron density of reference GaAs epilayers grown at 580°C with the same GaTe beam-equivalent pressure of each NW series. The doping level of the reference GaAs layers was evaluated by means of Hall measurements at room temperature. Three NW series have been grown for reference doping $n = 7 \; 10^{17}$ cm$^{-3}$(series A), $4 \; 10^{18}$ cm$^{-3}$(B), and $1 \; 10^{19}$ cm$^{-3}$ (C), respectively. The fourth series (U) consists of nominally undoped NWs. Unintentionally doped reference layers, grown right before the undoped NWs, demonstrated a background *p*-carrier density of $5 10^{15}$cm$^{-3}$. A summary of the samples used for this work can be found in **Table 1**.

**Table 1.** Growth conditions of all the GaAs NWs used in this work. $T_g$ is the NWs growth temperature. Reference layer doping is the Hall carrier density measured on epitaxial layers grown at 580° C with the same GaTe flux corresponding to the relative NW samples. Growth duration was 60 min for all samples, except for C610b (75 min).

| Series | Reference layer doping [cm$^{-3}$] | NWs $T_g = 580°C$ | NWs $T_g = 610°C$ | NWs $T_g = 640°C$ |
|---|---|---|---|---|
| U | p = 5x10$^{15}$ | U580 | U610 | U640 |
| A | n = 7x10$^{17}$ | A580 | A610 | A640 |
| B | n = 4x10$^{18}$ | B580 | B610 | B640 |
| C | n = 1x10$^{19}$ | C580 | C610a,C610b | C640 |

Representative secondary electron microscopy (SEM) images of NWs grown for 60 min at different substrate temperatures with increasing GaTe flux are shown in **Figure 1**(a-f). NWs were found to form by self-assisted epitaxy at all of the growth temperatures examined with a growth rate of 6.5±0.5 µm/h. Temperature-dependent or doping-dependent differences in the growth rate were not observed within the quoted experimental uncertainty, which reflects the statistical distribution of the NW length as determined from the SEM images.

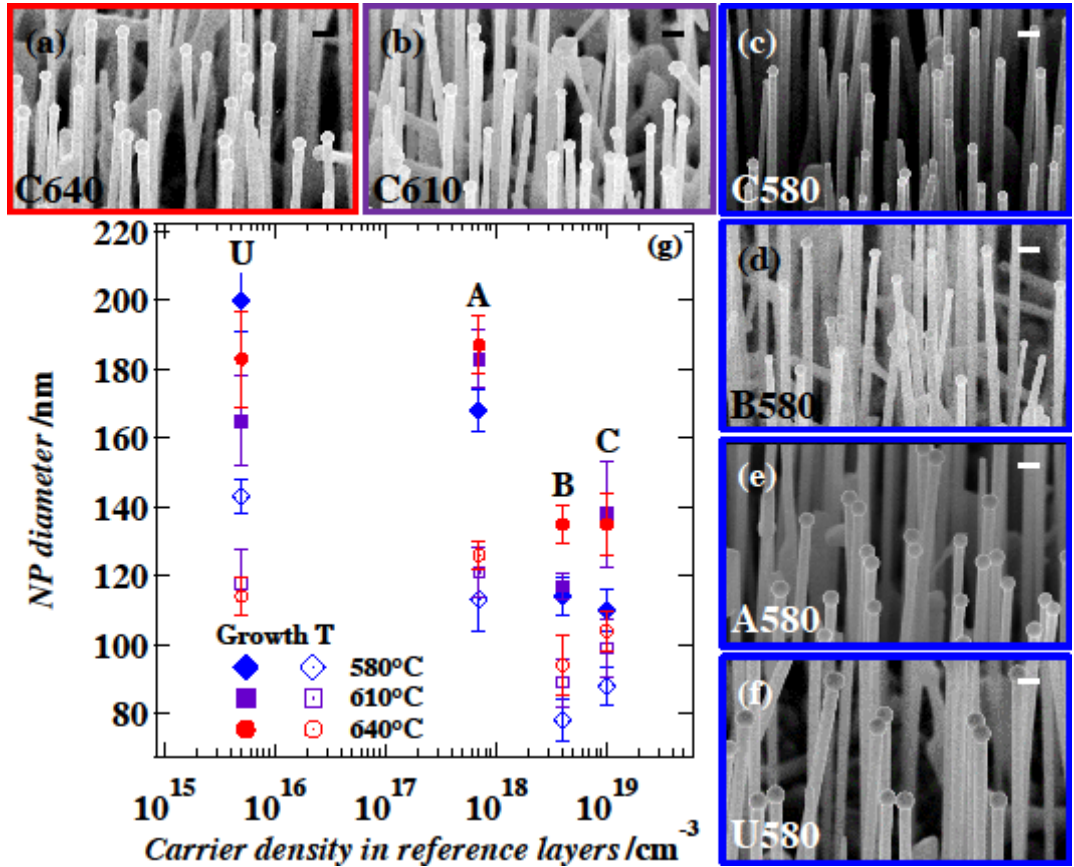

**Figure 1.** (a) to (f) Tilted-view SEM images of samples grown at different temperature and GaTe fluxes as indicated in the figure. The horizontal markers correspond to 200 nm. (g) Diameter of the nanowires (open symbols) and of the Ga nanoparticle at the nanowire tip (solid symbols) as a function of the carrier density measured on the calibration epitaxial layers.

In Figure 1(g) we report the diameter of the Ga nanoparticle (NP), and of the NWs in the region close to the NP, as a function of the reference doping at different $T_g$. The data correspond to average values on more than twenty NWs measured on SEM images such as those in Fig. 1(a-f). In standard conditions of self-assisted growth, the Ga NP is a quasi-complete sphere with a typical contact angle of 130° with the NW body.[21] We observe that co-evaporation of Te induces a reduction of NW and NP diameter. This effect is more pronounced in case of growth at the lowest temperature of 580 °C, where the thickest undoped NWs are obtained and the highest Te incorporation in the NWs is expected. The average diameter of the NPs (NWs) decreases from 200 nm (140 nm) [Figure 1(f)] in the undoped case to 170 nm (110 nm) already for the lowest doping level [Figure 1(e)]. We will see that this feature reflects a non-negligible incorporation of Te in the NWs, as detected by PL. By increasing the GaTe flux, the NP (NW) diameter shrinks to 110 nm (90 nm), and some tapering of NWs is observed (see panel (c)). A similar behavior is also found at higher growth temperature, although on a reduced scale. At $T_g$ = 640°C, A and U wires have nearly the same NP (and NW) average diameter of 180 nm (120 nm). Increasing the Te flux it decreases to 130 nm (100 nm). The slower Te-induced reduction of the diameter at the higher growth temperature well correlates with the PL data, as it will be detailed in the following.

Another effect induced by Te on NW morphology is the presence of lateral parasitic growth ("stems") that can be noticed in the circles in **Figure 2**. This parasitic growth increases with the GaTe flux and is observed also in the samples grown at higher temperature.

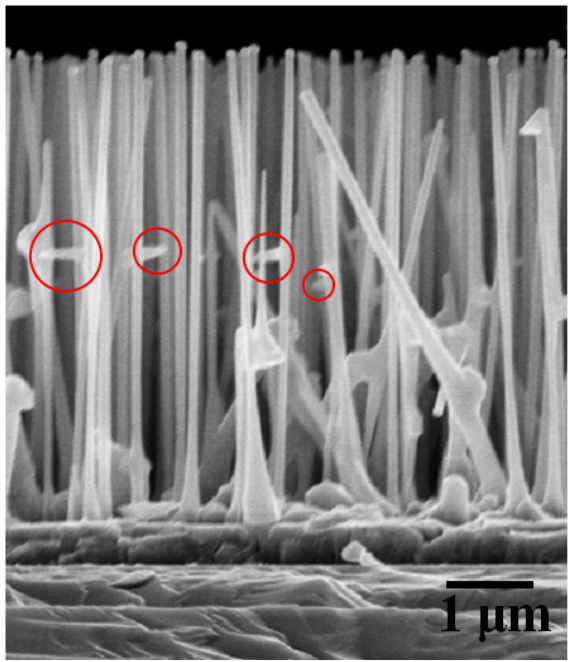

**Figure 2.** SEM image of sample C580 observed in cross-section on a cleaved edge of the as grown sample. The presence of lateral stems is visible in the red circles.

The presence of Te is known to influence in a complex way the surface kinetics during epitaxial growth of Te-doped III-V layers (see Ref. 22 and refs therein). Czaban et al.[7] evidenced its impact on the growth of *n-p* core/shell GaAs NWs by gas source MBE. Enhanced radial growth at the expense of axial growth was observed as a function of the growth time of the Te-doped core. This was ascribed to Te-induced hindering of Ga diffusion on the substrate and on the NW sidewalls.

The low As/Ga flux ratio employed in the present work allowed a sufficient Ga inflow at the NW tips to maintain a vapor-liquid solid growth at all growth conditions, as demonstrated by the presence of the Ga NPs at the NW tips. The presence of Te only induced the diameter reduction described above in Figure 1. The reduced Ga mobility, however, may be responsible also for the nucleation of Ga NPs on the NW lateral surface, assisting the growth of stems shown in Figure2.

## 3. Charge transport analysis

Three samples expected to exhibit a sufficient conductance (B580, C580 and C610) were investigated at the level of the individual nanostructure by fabricating single-NW FET devices and performing transport measurement on them. Several devices were fabricated and measured from each grown sample (see Section 7 for details about the device architecture).

As anticipated in the introduction, Hall effect is very hard to implement on NWs and few demonstration exist so far.[14,15,16] Apart from further methods based on the Seebeck response,[18] the most convenient and accessible way to estimate the carrier density and mobility in a NW is based on its field-effect transconductance.[23,24] Indeed, as long as the device works in a linear regime of charge accumulation, the device conductance $G$ is related to the mobility $\mu$ through the relation

$$G(V_{bg}) = \frac{\mu C_{bg}}{L^2}(V_{bg} - V_{th})$$

where $C_{bg}$ is the capacitance between the backgate and the NW, $V_{bg}$ is the backgate voltage $L$ is the length of the NW between the source and drain contacts, and $V_{th}$ is the threshold voltage of the device. In the linear regime, mobility can be directly extracted from the device transconductance $g_m = dG/dV_{bg}$. Based on the value of $\mu$, the carrier density can be calculated starting from the device resistance and the physical dimensions of the FET channel: this procedure is equivalent to a linear extrapolation of $V_{th}$ and to a calculation of the carrier density following the equation

$$n = \frac{C_{bg}V_{th}}{eL\pi r^2}$$

where $r$ is the NW radius and $e$ is the elementary charge.

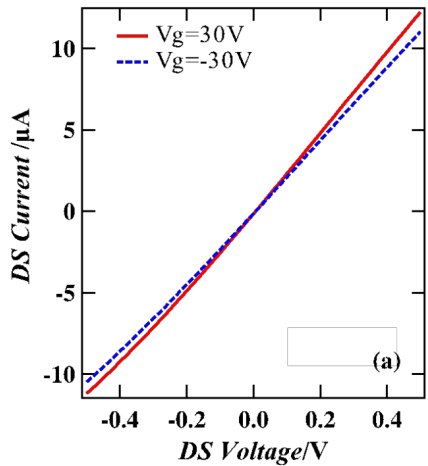

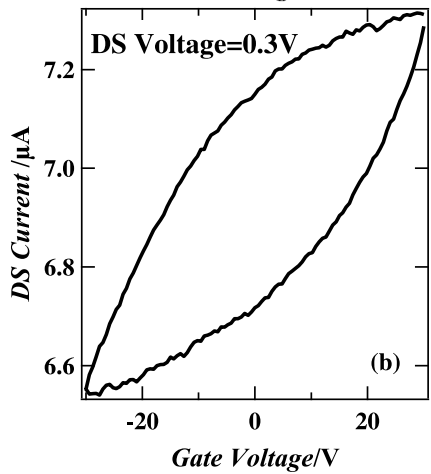

**Figure 3.** Characteristic curves of a NW-FET device based on a single Te-doped GaAs NW from sample C580, as measured at room temperature in air. In (a) the behavior of I-V curves as a function of back-gate bias indicates n-type doping. In (b) the transconductance characteristic is shown.

In **Figure 3**(a) we report I-V curves as a function of the gate voltage for one of the studied device from sample C580. The corresponding transconductance curves are visible in Figure 3(b). A large hysteresis is observed upon sweeping the gate: this effect is due to the screening by surface states on the NWs, whose population depends on the gating history of the device.[18,23] The doping density was estimated from the evolution of the I-V characteristics as a function of the back-gate voltage, following the model previously discussed. The value of $C_{bg}$ was estimated using the model by Wunnicke[26] that assimilates the NW to a metallic cylinder on an infinite metal plate:

$$C_{bg} = L \frac{2\pi\varepsilon}{\cosh^{-1}(h/r)}$$

where $h$ is the sum of the oxide thickness and the NW radius $r$ and $\varepsilon = \varepsilon_0\varepsilon_r$ is the oxide dielectric constant. An effective relative dielectric constant of $\varepsilon_r = 2.25$ instead of 3.9 was used to take into account the hexagonal section of the NWs and the fact that the NWs are not embedded in the oxide.[26] The NW diameter and the length of the NW between the source and drain contacts was measured from SEM images of individual NW-FET acquired after the electrical measurements.

Due to the large doping level found in the measured GaAs NWs, the threshold voltage could not be directly measured and had to be extrapolated from the transconductance in the available gating range. In particular the values of the two sweeps (as for instance in Figure 3(b)) were averaged point by point to exclude the hysteresis effect and then extrapolation was performed assuming a linear dependence of the conductance as a function of the gate bias. The resulting transport parameters are reported in **Figure 4**.

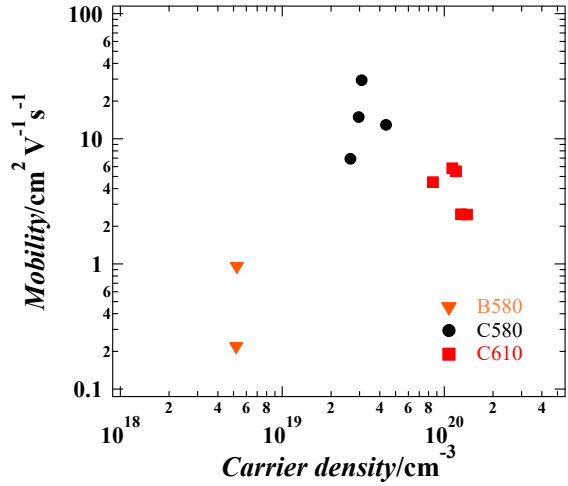

**Figure 4.** Mobility and doping density as estimated in single wire NW-FET devices on samples B580, C580, C610(a and b).

Very large carrier density is estimated for all the measured samples as a consequence of the successful incorporation of Te in the NW structure. Moreover, the estimated carrier densities have little dispersion within the device ensembles. On the other hand, the approximations behind the field-effect method call for a discussion about possible sources of errors in the extracted parameters. Field-effect estimates for carrier density are, first of all, known to lead to a systematic overestimate of the carrier density and underestimate of mobility due to screening effects by surface states.[14,18] In addition, the presence of non-negligible contact resistances can affect both $n$ and $\mu$ estimates and lead in particular to an incorrect extrapolation of the threshold voltage $V_{th}$.[24] This is particularly true in the case of GaAs NW, where the small contact area and surface depletion make it generally hard to obtain reliable contacts. In our case, thanks to a combined study of the devices using Kelvin probe microscopy (Section 4), the voltage drop could be directly measured on selected devices and it was found to account to about ten per cent of the total bias, corresponding to a resistance of a few kΩ. This leads to an overestimation by a factor of 1.1 of the carrier concentration. Assuming a rather conservative residual 50% of the voltage drop over the contacts, this is expected to lead to an overestimation by a factor 2 of the threshold voltage.

From the data in Figure 4 we observe that the growth temperature does not play a major role in determining the carrier density, at least for high dopant flux. Te is known to give carrier density as high as 2 $10^{19}$cm$^{-3}$ in GaAs without compensation effects.[6] Our results support the same conclusion in the case of NWs. Indeed, the carrier density measured in our NW-FETs exceeds the doping density measured for the calibration epilayers by a factor that goes from 1 to 3 times at the lowest growth temperature, up to one order of magnitude at 610°C. Despite possible systematic errors in a simple field-effect analysis, our results clearly indicate that the incorporation mechanism in NWs is more effective than in 2D epitaxy. It is worth noting that such a high efficiency has been recently reported for GaN NWs doped with silicon.[28]

A higher doping efficiency in NWs can be ascribed to the different growth mechanisms occurring in wires and epilayers and, in particular, to different dopant incorporation paths that in NWs include both the diffusion through the NP and the incorporation at the sidewall facets. Moreover, in our specific case, the relatively low As/Ga beam pressure ratio used for NW growth, as compared to standard GaAs epitaxy, might in fact reduce the competition between As and Te incorporation at the As sites, reducing the Te re-evaporation and allowing effective doping at growth temperature considerably higher than in case of 2D MBE growth.

The estimated mobility values display, within each group of nominally identical devices, a larger scattering. This is a likely consequence of the intrinsic limits of the field-effect method. Indeed, as already mentioned, our estimates do not take into account the contact resistance, which typically displays a sizable variability in the case of GaAs due to presence of residual Schottky barriers. We also note that, surprisingly, mobility estimates for sample B580 are sizably lower, despite the lower doping level. We interpret this effect as due to contact resistances, which are expected to rapidly increase if the doping level decreases and can lead to a strong underestimation of mobility.

A comparison of our results with data from literature is made difficult because quantitative analysis about carrier density and mobility in $n$-doped III-V NWs is scarce. High doping levels have also been obtained in InP:Sn NWs, with electron concentration in the $10^{19}$ cm$^{-3}$ range.[16] However no value of the carrier mobility has been provided in that work. Mobility as high as 2200 cm$^2$ V$^{-1}$ s$^{-1}$ has been reported in modulation doped GaAs/AlGaAs NWs,[29] but with carrier density in the $10^{16}$ cm$^{-3}$ range, three orders of magnitude lower than in our case.

## 4. Kelvin probe analysis

In order to obtain mobility data that are not affected by the contacts properties, individual NWs were measured by Kelvin probe force microscopy (KPFM).[30] By measuring the surface potential profile on a biased nanowire, KPFM can be used as a "four-point-probe" where the scanning tip replaces the two inner contacts of the conventional setup.[31] KPFM allows the quantitative determination of surface potential of semiconductor with nanometer resolution and have been widely used for the study of semiconductor NWs.[31-33] Additionally, at the heavy doping level of the measured GaAs NWs, surface potential profiles are a good approximation of their bulk potential profiles. Therefore, the potential profiles were used here to evaluate independently the NW carrier mobility.

KPFM measurements were performed on the GaAs NWs from sample C610, where the highest carrier density was observed (see Figure 4), in FET device configuration, where source and back-gate contacts were grounded and the drain contact was biased. The topography and the surface potential map of a device measured under a 1 V drain bias are shown in **Figure 5**(a) and (b), respectively. The surface potential profiles along the axial direction of the NW for a bias range from -1 V to +1 V with 0.5 V steps are shown in Figure 5(c). The difference between the surface potential measured at the drain contact with bias -1 and +1 V is of 1.66 V instead of 2 V, indicating that part of the applied potential drops outside the NW device. Moreover, even if the current-voltage $I_{ds}$-$V_{ds}$ curve in the range of the voltage applied has a linear behavior (not shown), the surface potential profiles show a potential drop at both contacts consistent with the presence of residual Schottky barriers.[34] Both the above mentioned effects however have no impact on the measure of the actual voltage drop across the NW, because the extraction of the slope from the surface potential profile in the central part of the NW gives direct access to the measure of the electric field across it. This allows a better evaluation of the mobility that can be estimated as follows. At a given applied bias the mobility $\mu_n$ is given by:

$$\mu_n = \frac{I}{qA(x)N_D \frac{d\varphi}{dx}}$$

where *I* is the current, *A(x)* is the NW cross section and $\frac{d\varphi}{dx}$ is the derivative of the measured surface potential with respect to the position along the NW axis as calculated in the linear portion of the *ϕ(x)* curve. *A(x)* is extracted from the simultaneous AFM topographic measurement. For these growth conditions, field effect transport measurements (see Figure 4) provide a carrier density *n* = 1.1 ± 0.2 x $10^{20}$ cm$^{-3}$. At this value, KPFM yields a carrier mobility $\mu_n$ = 21.6 ± 0.4 cm$^2$/Vs, larger than the value given by I-V curves. This confirms that a simple analysis of I-V curves taken in FET mode, which neglects the contact resistance, leads to an underestimation of the carrier mobility, as discussed above, while KPFM allows a better estimation by measuring the actual voltage drop across the NWs.

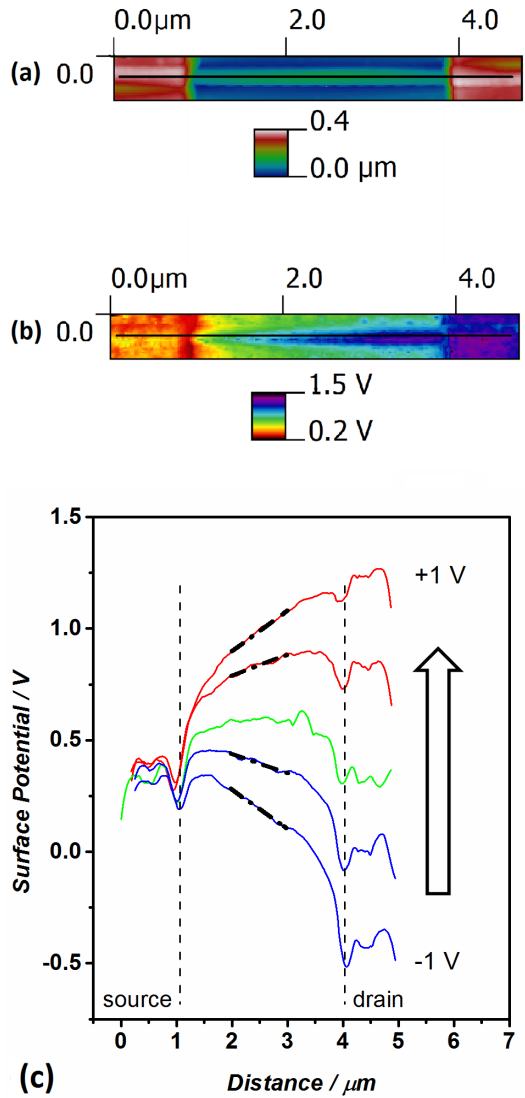

**Figure 5.** (a) Measured topography and (b) surface potential of a Te-doped GaAs NW device fabricated on a NW from sample C610 under a bias of 1 V applied to the drain contact. The black lines along the NW in (a) and (b) show the position where the surface potential profiles were extracted. (c) Measured surface potential profiles taken along the axis of the same Te-doped GaAs NW at several drain biases, spanning from −1 to +1 V with 0.5 V steps. The location of the source and drain pads are marked by dashed lines.

## 5. Photoluminescence

Incorporation of Te was investigated by low temperature PL of all as-grown samples. For sake of comparison with the PL spectra of the Te doped NWs, we first report in **Figure 6** the PL spectra of the undoped NWs and of the doped 2D layers. Figure 6 (a) shows the PL spectra of samples U580, U610 and U640, while Figure 6 (b) shows the spectrum of the reference Te-doped GaAs epilayer grown with the highest doping level (*n* = 1 $10^{19}$ cm$^{-3}$). Spectra in Figure 6(a) are shown in logarithmic scale to point out the absence of any radiative recombination at energies higher than that of the excitonic recombination (at about 1.516 eV). The PL is typical for our self-catalyzed GaAs NWs that have zincblende lattice structure.[36] The large band at low energy in Figure 6(a) is probably due to residual carbon incorporation, which decreases at higher Tg as pointed out by the corresponding increase of the relative intensity of the excitonic recombination with respect to the broad, impurity-related band. The spectrum of the doped epilayers (Figure 6(b)), compared with the calibrated spectra reported by DeSheng and coworkers,[6] suggests that the doping level is about $8\ 10^{18}$ cm$^{-3}$, in very good agreement with the value measured by Hall measurements ($1.0\ 10^{19}$ cm$^{-3}$). The narrower peaks around 1.5 eV are luminescence signal from the undoped buffer layer grown beneath the doped layer (see Section 7). Notice the shoulder at 1.65 eV, Similarly, although a detailed electrical characterization of our NWs was not possible for low doping levels, we can exploit the relationship between PL and electrical measurements to gain semi-quantitative information on the doping level of our samples as a function of the growth parameters using a qualitative analysis of the PL line-shape.

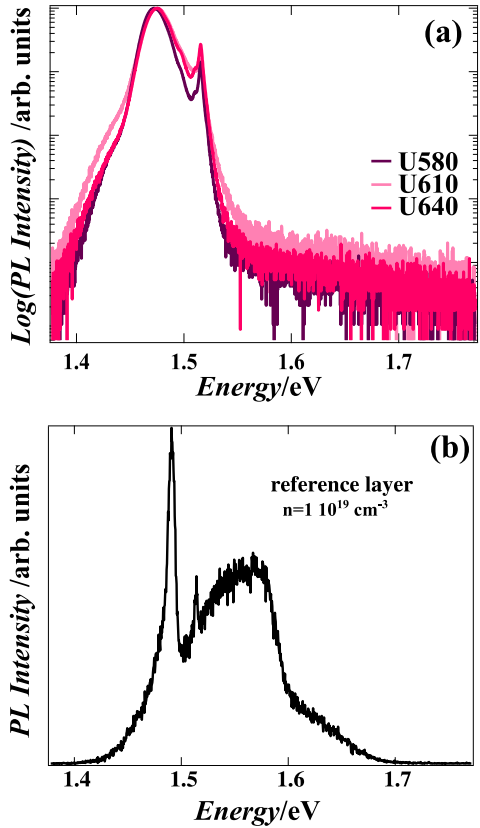

**Figure 6.** (a) Logaritmic PL intensity from the undoped NW samples, U580, U610, and U640 The logarithmic scale points out the absence of any PL signal at energies higher than that of the free-exciton. (b) PL of the reference layer with the highest doping level. Moss-Burstein shift of the luminescence is clearly observed.

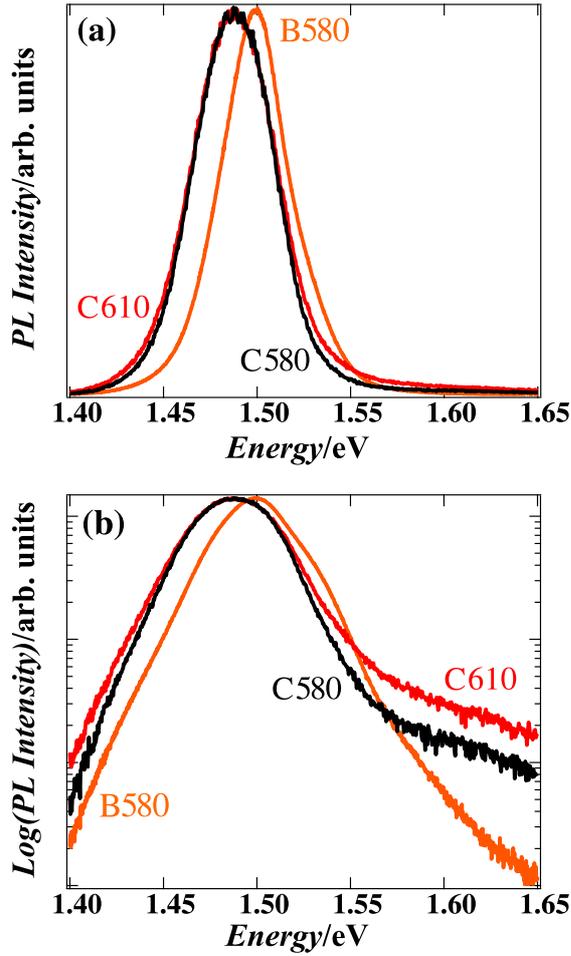

**Figure 7.** Low temperature (14K) photoluminescence spectra measured for samples B580, C580 and C610 plotted using a linear (a) and a logarithmic scale (b). Spectra intensities are normalized to unity.

In **Figure 7** we show in linear (a) and logarithmic (b) scale the PL spectra of the GaAs NWs samples used for transport measurements: B580, C580 and C610. The linear spectra (Figure 7(a)) present a single, broad main peak that shows a red shift as the carrier density increases. At the carrier densities determined by transport measurements and reported in Figure 4, strong many-body effects are expected in GaAs,[34] which should cause, in particular, a band-gap reduction (band-gap renormalization). At the same time PL may provide evidence for the Moss-Burstein (MB) blue-shift. The position of the PL peak will be given by the competition of these effects that depend on carrier density, carrier masses and joint density of states. In particular, the spectra in Figure 7(a) show a red-shift of the PL onset at low energy that can be interpreted as a clear evidence for band-gap renormalization, while no evidence for MB shift is observed. Our spectra, similar to those obtained in p-type Sn-doped GaAs NWs,[37] show remarkable differences with those obtained in epitaxial layers (see Figure 6(b) and Ref. 6) that, apart for the excitonic lines of the buffer layer, possess a strong broad band with a maximum well above the GaAs band gap, a clear indication of a MB shift. The maximum of the broad band of the NWs luminescence lies below the exciton energy, a clear indication that the band shrinking due to many-body effects is stronger than the MB shift. We notice that in other NWs systems (InP:Sn[16] and GaAs:Zn[38]) a stronger MB shift is observed, in closer resemblance with epilayer results (Fig. 6 and Ref. 6). However, if we look at the logarithmic plot of the luminescence, we clearly see features at high energies that are not observed in undoped samples (see Figure 6(a)). In particular samples C580 and C610, those with the highest carrier density in transport measurements, show a clear shoulder, similar to that observed in our epitaxial layers at the highest doping levels. Subtracting the background contribution, the maximum of this recombination band is found at about 1.65 eV, as in the epitaxial layers, see Fig. 6(b). Although in a qualitative way, PL in the doped NWs is then in agreement with transport result, showing stronger effects as the doping increases.

We can then use PL data to obtain information about the lower doping levels, obtained with lower GaTe fluxes for which the transport measurements cannot be performed. In **Figure 8** we show PL spectra of samples A610 (a) and A580 (b). Comparing the spectra of A610 with those for undoped NWs in Figure 6 (a) we do not observe major differences: the presence of the excitonic recombination suggests that the Te incorporation in A610 is of the order of the background doping.

Remarkable PL effects that indicate Te incorporation as an electrical active impurity are instead observed in sample A580, Figure 8(b): The excitonic line has disappeared and only a single broad band is observed, that blue-shifts as the optical excitation intensity increases.

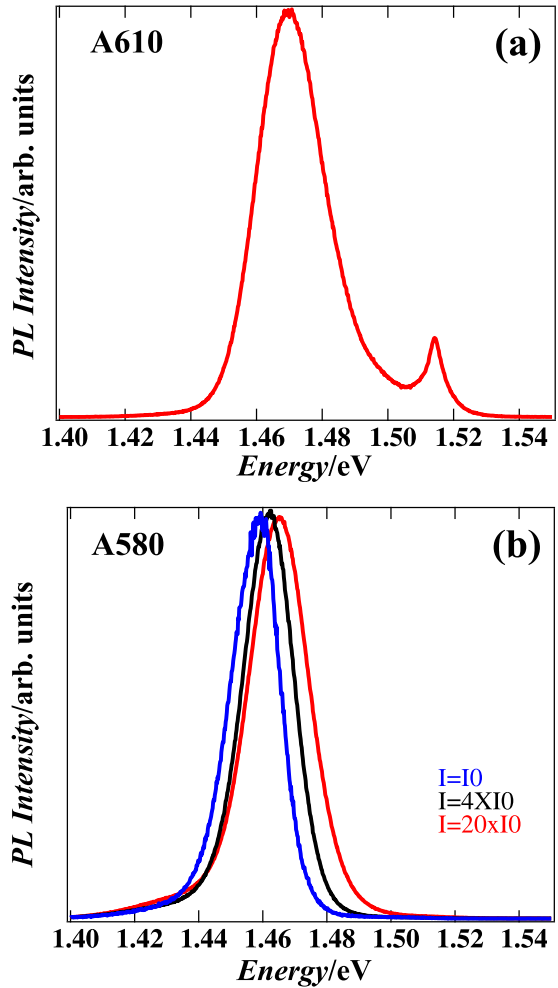

**Figure 8.** PL measurements at 14 K of samples A610 (a) and A580 (b). The PL of A610 shows excitonic recombination, a clear sign of a low impurity level in sample A610, comparable with that of the undoped samples. The blue-shift of the luminescence of sample A580 for increasing excitation intensity is compatible with a donor-acceptor pair recombination due to the presence of measurable levels of Te incorporation.

This behavior is not observed for the broad band in sample A610. Moreover the single band of sample A580 is broader than the broad band of A610. The blue shift for increasing excitation powers allows us to attribute the broad band of A580 to donor-acceptor pair (DAP) recombination, Te being the donor and residual carbon the acceptor. This typical DAP behavior, as expected, is not observed for the highly doped samples shown in Figure 7.

The data in Figure 8 indicate that at moderate dopant flux, Te incorporation strongly depends on the growth temperature, being detectable only in sample A580, grown at the lowest $T_g$. On the other hand, both electrical and optical measurements suggest a slightly higher doping level for sample C610 than for C580, that has been grown at a lower temperature that should give rise to a lower re-evaporation rate of Te. A tentative explanation of this behavior could be given considering also As evaporation. The high Te flux used for those samples probably saturates any Te re-evaporation at both growth temperatures. At the highest among the two, however, a small increase of As re-evaporation could favor a slightly higher Te incorporation as donors in the As empty sites.

Undoped GaAs NWs grown with our method have zincblende structure.[36] Suomalainen and coworkers[16] give some hint that the incorporation of Te impurities provokes the formation of wurtzite sections along the wires. In this case the intensity-dependent behavior shown Fig. 8(b) would be compatible with the observation of spatially indirect recombination in the type II aligned ZB-WZ heterostructure.[39] However we point out that in the GaAs NWs grown in our laboratory and showing the contemporary presence WZ and ZB, a sharp peak has always been observed in the excitonic energy region.[40,41] In our Te-doped samples, instead, no excitonic recombination is longer observed, an indication that the PL changes should be attributed to the incorporation of tellurium. Moreover, in the samples with higher doping, the presence of recombination at energies above the band gap of ZB is the clear indication that no type-II transition is implied.

## 6. Conclusion

The ensemble of experiments performed on GaAs NWs demonstrates that Te, obtained from a GaTe source, can be reliably used to achieve high free electron density in GaAs NWs grown by self-assisted MBE. An effective incorporation of Te can be obtained in NWs also at growth temperatures considerably higher than those used for Te doping in 2D MBE, as long as a significant flux of Te atoms, is provided. This feature allows the optimization of the NW morphology, as well as reducing the presence of unwanted impurities, while maintaining a high carrier density.

The combined use of single wire transport measurements, KPFM and PL allowed us to produce a detailed characterization of the incorporation of dopant impurities in a very wide range of concentration, and correlate in a semi-quantitative way Te incorporation with the growth conditions. The combined data provided in this work then offer to the NW community a reliable data basis to access via PL to a quick quantitative determination of the Te doping level of GaAs NWs, without the necessary, complex processing to achieve ohmic contacts. Finally, we point out that our work has been performed on NWs produced by self-induced growth, i.e. without the use of external elements to induce one-dimensional growth, a feature that would permit the integration of our doped NWs in the Si technology platform.

## 7. Experimental Section

*Nw growth*: GaAsNWs were grown on Si-treated[36] n-type GaAs (111)B wafers by solid source MBE using a As/Ga beam equivalent pressure ratio (BPR) of 5 and a Ga flux corresponding to a two dimensional growth rate of 1 µm/h. A Riber standard Knudsen cell operating in the 100-150 °C range, loaded with 1 g of GaTe from Fox Chemicals GmbH was used as Te source. The substrate temperature was varied between 580 and 640°C and measured by means of an infrared pyrometer. Growth time was typically 60 minutes. The sample investigated by KPFM (C610b) was grown for 75 minutes. Four series of NW samples were grown varying the Te flux obtained from a GaTe Knudsen cell. The series are indexed according to the carrier density measured in reference GaAs epilayers of 1µm thickness grown on semi-insulating GaAs(100) substrates at 580°C at BPR=10 with a growth rate of 1 µm/h with the same Te fluxes. A 0.5 µm thick undoped buffer layer was grown in these reference samples beneath the doped layer. The carrier density of the epitaxial layers was evaluated by Hall measurements at room temperature in the van der Pauw configuration.

*Transport measurements:* Doping levels of the NWs were measured using single-NW-based field-effect transistors (NW-FET) in the back gate geometry. GaAs NWs were dispersed in isopropyl alcohol by sonication of the as grown substrate and the solution was drop-casted on top of a pre-patterned degenerately doped Si substrate covered by a layer of 280 nm of $SiO_2$. The NWs position was determined by SEM imaging and contact electrodes were fabricated by e-beam lithography and evaporation of a Pd/Ge/Au (50/150/60 nm) multilayer. Prior to evaporation, resist residues were removed with and oxygen plasma and the exposed GaAs NW surface was treated with a $HCl:H_2O$ (1:4) solution for 1 minute to remove the native oxide. Thermal annealing at 290°C for 30 seconds was used to improve the conductivity of the ohmic contacts to the NWs. A typical device is depicted in the SEM image in **Figure 9**.

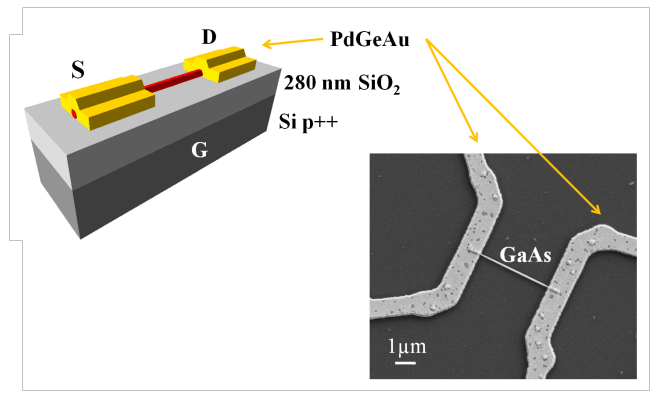

**Figure 9.** (a) Schematic of the single wire devices used for transport measurements and KPFM analysis. (b) SEM image of a single wire device.

*Kelvin Probe force microscopy:* KPFM was performed by a commercial system (Dimension Edge, Bruker) inside a dry nitrogen glove box. The surface potential was measured simultaneously with the topographic signal using amplitude modulation KPFM at an effective tip sample distance of 5–10 nm during scanning. The topographic height was obtained by maintaining the amplitude of the first cantilever resonance ($f_{1st} \cong 75$ kHz) at a predefined amplitude set point of 5 nm. The surface potential (*SP*) was determined by compensating the ac component of the electrostatic force, $F_{ES}$, with an applied dc voltage (= |*SP*|) in a feedback control loop. To separate topographic from *SP* signal, increase the sensitivity, and minimize probe convolution effects, the ac electrostatic force component was generated at the second resonance of the cantilever,[42] $f_{2nd} \cong 450$ kHz, by applying an ac voltage of about 500 mV.

*Photoluminescence:* PL was excited by the 514.5 nm line of an Ar$^+$ laser, dispersed by a 1 m long monochromator and detected in single-photon counting mode. Measurements were performed on as grown samples at *T*=14 K.

**Acknowledgements**
MO acknowledges the financial support from Area Science Park in Trieste. S. Roddaro acknowledges the support of CNR through the bilateral CNR-RFBR project 2015–2017.

Received: ((will be filled in by the editorial staff))
Revised: ((will be filled in by the editorial staff))
Published online: ((will be filled in by the editorial staff))

[1] M. Piccin, G. Bais, V. Grillo, F. Jabeen, S. De Franceschi, E. Carlino, M. Lazzarino, F. Romanato, L. Businaro, S. Rubini, F. Martelli, A Franciosi, *Phys. E Low-dimensional Syst. Nanostructures* **2007**, *37*, 134.

[2] C. Colombo, M. Heiβ, M. Grätzel, A Fontcuberta i Morral, *Appl. Phys. Lett.* **2009**, *94*, 173108.

[3] M. Hilse, M. Ramsteiner, S. Breuer, L. Geelhaar, H. Riechert, *Appl. Phys. Lett.* **2010**, *96*, 193104.

[4] A. D. Mohite, D. E. Perea, S. Singh, S. A. Dayeh, I. H. Campbell, S. T. Picraux, H. Htoon, *Nano Lett.* **2012**, *12*, 1965.

[5] M. D. Brubaker, P. T. Blanchard, J. B. Schlager, A. W. Sanders, A. M. Herrero, A. Roshko, S. M. Duff, T. E. Harvey, V. M. Bright, N. A. Sanford, K. A. Bertness, *J. Electron. Mater.* **2013**, *42*, 868.

[6] J. De-Sheng, Y. Makita, K. Ploog, and H. J. Queisser, *J. Appl. Phys.* **1982**, *53*, 999.

[7] J. Czaban, D. A. Thompson, R. R. LaPierre, *Nano Lett.* **2009**, *9*, 148.

[8] J. Caram, C. Sandoval, M. Tirado, D. Comedi, J. Czaban, D. A. Thompson, R. R. LaPierre, *Nanotechnology* **2010**, *21*, 134007.

[9] O. Salehzadeh, K. L. Kavanagh, S. P. Watkins, *J. Appl. Phys.* **2012**, *112*, 054324.

[10] A. Darbandi, O. Salehzadeh, P. Kuyanov, R. R. LaPierre, S. P. Watkins, *J. Appl. Phys.* **2014**, *115*, 234305.

[11] S. Suomalainen, T. V. Hakkarainen, T. Salminen, R. Koskinen, M. Honkanen, E. Luna, M. Guina, *Appl. Phys. Lett.* **2015**, *107*, 012101.

[12] C. Gutsche, I. Regolin, K. Blekker, A. Lysov, W. Prost, F. J. Tegude, *J. Appl. Phys.* **2009**, *105*, 024305.

[13] C. Gutsche, A. Lysov, I. Regolin, K. Blekker, W. Prost, F. J Tegude, *Nanoscale Res. Lett.* **2011**, *6*, 65.

[14] K. Storm, F. Halvardsson, M. Heurlin, D. Lindgren, A. Gustafsson, P. M. Wu, B. Monemar, L. Samuelson, *Nat. Nanotechnol.* **2012**, *7*, 718.

[15] Ch. Blömers, T. Grap, M. I. Lepsa, J. Moers, St. Trellenkamp, D. Grützmacher, H. Lüth, T. Schäpers, *Appl. Phys. Lett.* **2012**, *101,* 152106.

[16] D. Lindgren, O. Hultin, M. Heurlin, K. Storm, M. T. Borgström, L. Samuelson, A. Gustafsson, *Nanotechnology* **2015**, *26*, 045705.

[17] A. Pitanti, S. Roddaro, M. S. Vitiello, A. Tredicucci, *J. Appl. Phys.* **2012**, *111, 064301*.

[18] S. Roddaro, D. Ercolani, M. A. Safeen, S. Suomalainen, F. Rossella, F, Giazotto, L. Sorba, F. Beltram, *Nano Lett.* **2013**, *13*, 3638.

[19] W.N. Jiang, N. X. Nguyen, R. D. Underwood, U. K. Mishra, R. G. Wilson, *Appl. Phys. Lett.***1995**, *66*, 845.

[20] X. Gan, X. Zheng, Y. Wu, S. Lu, H. Yang, M. Arimochi, T. Watanabe, M. Ikeda, I. Nomachi, H. Yoshida, S. Uchida, *Jpn. J. Appl. Phys.* **2014**, *53*, 021201.

[21] T. Rieger, S. Heiderich, S. Lenk, M. I. Lepsa, D. Grützmacher, *J. Cryst. Growth* **2012**, *353*, 39.

[22] B. Paquette, B. Ilahi,V. Aimez, R. Arès, *J. Cryst. Growth* **2013**, *383*, 30.

[23] X. Jiang, Q. Xiong, S. Nam, F. Qian, Y. Li, C. M. Lieber, *Nano Lett.* **2007,** *7*, 3214.

[24] Ö. Gül, D. J. van Woerkom, I. van Weperen, Diana Car, S. R. Plissard, E. P. A. M. Bakkers, L. P. Kouwenhoven, *Nanotechnology* **2015**, *26*, 215202.

[25] M. S. Fuhrer, B. M. Kim, T. Dürkop, T. Brintlinger, *Nano Lett.* **2002**, *2*, 755.

[26] O. Wunnicke, *Appl. Phys. Lett.* **2006**, *89,* 083102.

[27] R. Martel, T. Schmidt, H. R. Shea, T. Hertel, Ph. Avouris, *Appl. Phys. Lett.* **1998**, *73*, 2447.

[28] Z. Fang, E. Robin, E. Rozas-Jiménez, A. Cros, F. Donatini, N. Mollard, J. Pernot, B. Daudin, *Nano Lett.* **2015**,15, 6794.

[29]     J. L. Boland, S. Conesa-Boj, P. Parkinson, G. Tütüncüoglu, F. Matteini, D. Rüffer, A. Casadei, F. Amaduzzi, F. Jabeen, C. L. Davies, H. J. Joyce, L. M. Herz, A. Fontcuberta i Morral, M. B. Johnston, *Nano Lett.* **2015**, *15*, 1336.c

[30]     M. Nonnenmacher, M. P. O'Boyle, H. Wickramasinghe, *Appl. Phys. Lett*. **1991**, *58*, 2921.

[31]     E. Koren, Y. Rosenwaks, J. E. Allen, E. R. Hemesath, L. J. Lauhon, *Appl. Phys. Lett.* **2009**, *95*, 092105.

[32]     E. Koren, N. Berkovitch, Y. Rosenwaks, *Nano Lett.* **2010**, *10*, 1163.

[33]     E. Halpern, G. Cohen,  S. Gross, A. Henning, M. Matok, A. V. Kretinin, H. Shtrikman, Y. Rosenwaks *Phys. Status Solidi A* **2014**, *211*, 473.

[34]     M. Orrù, V. Piazza, S. Rubini, S. Roddaro, *Phys. Rev. Appl.* **2015**, *4*, 044010.

[35]     J. C. Hensel, T.C. Phillips, C. A. Thomas, *Solid State Phys*. **1977,** *32*, 87.

[36]     S. Ambrosini, M. Fanetti, V. Grillo, A. Franciosi, S. Rubini, *J. Appl. Phys.* **2011**, *109*, 094306.

[37]     A. Lysov, M. Offer, C. Gutsche, I. Regolin, S. Topaloglu, M. Geller, W. Prost, F.-J. Tegude, *Nanotechnology* **2011**, *22*, 085702.

[38]     F. Yang, M.E. Messing, K. Mergenthaler, M. Ghasemi, J. Johansson, L.R. Wallenberg, Pistol, K. Deppert, L. Samuelson, M.H. Magnusson, *J. of Crystal. Growth* **2015**, *414*, 181.

 [39]     M. Murayama, T. Nakayama, *Phys*. *Rev*. *B* **1994**, *49*, 4710.

[40]     F. Martelli, M. Piccin, G. Bais, F. Jabeen, S. Ambrosini, S. Rubini, A. Franciosi, *Nanotechnology* **2007**, *18*, 125603.

[41]     M. De Luca, G. Lavenuta, A. Polimeni, S. Rubini, V. Grillo, F. Mura, A. Miriametro, M. Capizzi, F. Martelli, *Phys*. *Rev*. *B* **2013**, *87*, 235304.

[42]     A. Kikukawa, S. Hosaka, R. Imura,  *Rev. Sci. Instrum.* **1996,** *67*, 1463.

**Figure 1.** (a) to (f) Tilted-view SEM images of samples grown at different temperature and GaTe fluxes as indicated in the figure. The horizontal markers correspond to 200 nm. (g) Diameter of the nanowires (open symbols) and of the Ga nanoparticle at the nanowire tip (solid symbols) as a function of the carrier density measured on the calibration epitaxial layers.

**Figure 2.** SEM image of sample C580 observed in cross-section on a cleaved edge of the as grown sample. The presence of lateral stems is visible in the red circles.

**Figure 3.** Characteristic curves of a NW-FET device based on a single Te-doped GaAs NW from sample C580, as measured at room temperature in air. In (a) the behavior of I-V curves as a function of back-gate bias indicates n-type doping. In (b) the transconductance characteristic is shown.

**Figure 4.** Mobility and doping density as estimated in single wire NW- FET devices on samples B580, C580, C610(a and b).

**Figure 5.** (a) Measured topography and (b) surface potential of a Te-doped GaAs NW device fabricated on a NW from sample C610 under a bias of 1 V applied to the drain contact. The black lines along the NW in (a) and (b) show the position where the surface potential profiles were extracted. (c) Measured surface potential profiles taken along the axis of the same Te-doped GaAs NW at several drain biases, spanning from −1 to +1 V with 0.5 V steps. The location of the source and drain pads are marked by dashed lines.

**Figure 6.** (a) Logaritmic PL intensity from the undoped NW samples, U580, U610, and U640 The logarithmic scale points out the absence of any PL signal at energies higher than that of the free-exciton. (b) PL of the reference layer with the highest doping level. Moss-Burstein shift of the luminescence is clearly observed.

**Figure 7.** Low temperature (14K) photoluminescence spectra measured for samples B580, C580 and C610 plotted using a linear (a) and a logarithmic scale (b). Spectra intensities are normalized to unity.

**Figure 8.** PL measurements at 14 K of samples A610 (a) and A580 (b). The PL of A610 shows excitonic recombination, a clear sign of a low impurity level in sample A610, comparable with that of the undoped samples. The blue-shift of the luminescence of sample A580 for increasing excitation intensity is compatible with a donor-acceptor pair recombination due to the presence of measurable levels of Te incorporation.

**Figure 9.** (a) Schematic of the single wire devices used for transport measurements and KPFM analysis. (b) SEM image of a single wire device.

**Table 1.** Growth conditions of all the GaAs NWs used in this work. $T_g$ is the NWs growth temperature. Reference layer doping is the Hall carrier density measured on epitaxial layers grown at 580° C with the same GaTe flux corresponding to the relative NW samples. Growth duration was 60 min for all samples, except for C610b (75 min).

| Series | Reference layer doping [cm$^{-3}$] | NWs $T_g = 580°C$ | NWs $T_g = 610°C$ | NWs $T_g = 640°C$ |
|--------|-----------------------------------|-------------------|-------------------|-------------------|
| U | p = 5x10$^{15}$ | U580 | U610 | U640 |
| A | n = 7x10$^{17}$ | A580 | A610 | A640 |
| B | n = 4x10$^{18}$ | B580 | B610 | B640 |
| C | n = 1x10$^{19}$ | C580 | C610a, C610b | C640 |

**The control of doping in semiconductor nanowires** is a crucial issue in the design of reliable nanowire-based devices. The complementary use of transport measurements, Kelvin probe force microscopy and photoluminescence demonstrates efficiency and reliability of Te as donor impurity in GaAs nanowires grown by self-assisted molecular beam epitaxy.

Keywords: semiconductor nanowires, doping, carrier density

Marta Orrù, Eva Repiso, Stefania Carapezzi, Alex Henning, Stefano Roddaro, Alfonso Franciosi, Yossi Rosenwaks, Anna Cavallini, Faustino Martelli, and Silvia Rubini*

A Roadmap for Controlled and Efficient *n*-type Doping of Self-assisted GaAs Nanowires Grown by Molecular Beam Epitaxy

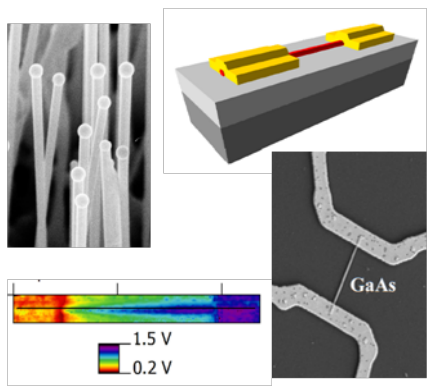